 \pdfoutput=1
\documentclass[%
preprint,
 amsmath,amssymb,
 aps,
floatfix,
]{revtex4-1}
\pdfoutput=1
\usepackage{appendix}
\usepackage{amsmath}
\usepackage{graphicx}
\usepackage{epstopdf}
\usepackage{graphics}
\usepackage{epsfig}
\usepackage[byname]{smartref}
\usepackage{hyperref} 
\usepackage{txfonts}
\usepackage{morefloats}
\usepackage{tocloft}
\usepackage{graphicx}
\usepackage{dcolumn}
\usepackage{bm}
\usepackage{hyperref}

\begin{document}

\preprint{V1}

\title{Spin Waves in 2D ferromagnetic square lattice stripe}

\author{Maher Z. Ahmed}
\affiliation{Department of Physics and Astronomy, University of Western Ontario, London ON N6A 3K7, Canada}
\affiliation{Physics Department, Faculty of Science, Ain Shams University, Abbsai, Cairo, Egypt}
\email{mahmed62@uwo.ca}
%



\begin{abstract}
In this work, the area and edges spin wave calculations were carried out
using the Heisenberg Hamiltonian and the tridiagonal method for
the 2D ferromagnetic square lattice stripe, where the SW modes are
characterized by a 1D in-plane wave vector $q_x$. The results show a general
and an unexpected feature that the area and edge spin waves only exist as
optic modes. This behavior is also seen in 2D Heisenberg antiferromagnetic
square lattice. This absence of the acoustic modes in the 2D square lattice
is explained in \cite{Selim2011} by the fact that the geometry constrains for NN
exchange inside the square lattice allow only optical modes. We suggest that
this unexpected behavior of spin waves in the 2D square lattice may be useful
in realizing an explanation for HTS.
\end{abstract}

\pacs{Valid PACS appear here}
\maketitle


\section{Introduction}
Spin waves in 2D magnetic systems are very interesting both experimentally
\cite{Katanin2011,PhysRevB.81.134409,PhysRevB.74.214424,Cervenka2009,Allenspach1994}
and theoretically
\cite{Arovas1988,JPSJ.58.3733,PhysRevB.39.2344,PhysRevB.60.1082,PhysRevB.40.2494,J.MiltonPereira2001,Pereira1999}.
For example, these systems are relevant to our understanding of high
temperature superconductors
\cite{Boothroyd2011,Dahm2009,RevModPhys.78.17,RevModPhys.66.763,Logvenov30102009,Guo2001,Hida1988},
and are the basis of many technological applications of ultrathin
ferromagnetic films (e.g., magnetic memory and storage devices, switches,
giant magnetoresistance, etc), as well as in the new promising field of
spintronics (see
\cite{DanielD.Stancil2009,Nguyen2007,ifmmode,Sarma2001,Mohn2006,Siegmann2006}).

Many theoretical techniques has been used to study  spin waves (SWs) in 2D
and 3D Heisenberg magnets \cite{PhysRevB.60.1082}. Some examples are the
Holstein-Primakoff (HP) method \cite{Selim2011}, the
``boson mean-field theory'' \cite{Arovas1988,JPSJ.58.3733} where Schwinger
bosons are used to represent the spin operators, and the ``modified spin-wave
theory'' \cite{PhysRevB.40.2494} where the Dyson-Maleev transformation is
used to represent the spin operators. Additionally, the semi-classical
approaches \cite{Kittel2005,Cottam2004} are widely employed for SWs at long
wavelengths (or small wavevectors). In recent years, ultra-thin magnetic
nanostructures have been fabricated and studied extensively
\cite{Nguyen2009,Nguyen2007,Hillebrands2002,Hillebrands2003,Hillebrands2006},
for their SW dynamics. It is often useful to distinguish between the
propagating SW modes and the localized SW modes. In a film geometry these are
usually referred to as "volume" (or "bulk") modes and "surface" modes
respectively. Thus, new theoretical studies are needed where the surfaces
(and/or interfaces) are important and where the localized SW modes are
considered in detail.

In this work our aim is to study SW modes at low temperatures (compared to
the Curie temperature $T_c$). We do this for ultra-thin ferromagnets with
typically one atom thickness, specifically a 2D stripe with a finite number
of atomic rows (a nanoribbon). We employ the Heisenberg model assume, for simplicity, a square lattice. We study both the
``area'' SWs (that propagate across the stripe width) and localized ``edge''
SWs. This work is very interesting for its expected novel fundamental physics
and promising application in magnetic devices.

An operator equation-of-motion technique known as the tridiagonal matrix
method \cite{CostaFilho2000195,Cottam1980,Cottam1976,PhysRev.185.720} will be
conveniently employed here to calculate the SW spectra and, in particular, to
distinguish between the edge modes and area modes of the ferromagnetic square
lattice nanoribbons.

\section{Theoretical model}
\begin{figure}[h]
\centering
\includegraphics[scale=0.5]{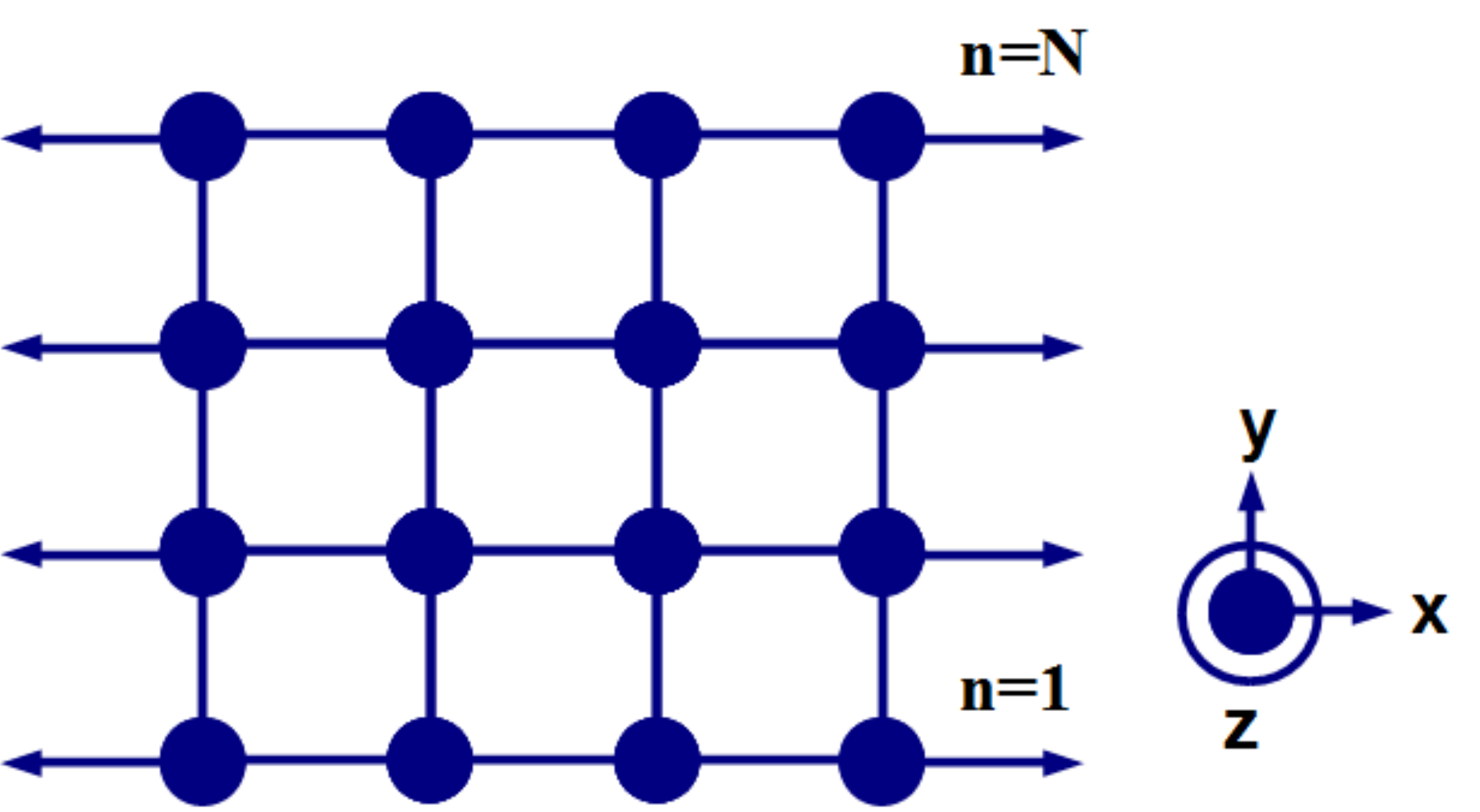}
\caption{Geometry of a 2D Heisenberg ferromagnetic square lattice nanoribbon. The spins are in the $xy$-plane and
the average spin alignment is in $z$ direction. The nanoribbon is finite in $y$ direction with $N$ atomic rows
($n = 1,\cdots,N$).}\label{fig:squarelattice}
\end{figure}

The system initially under study is a 2D Heisenberg ferromagnetic stripe (or
nanoribbon) in the $xy$-plane. We assume a square lattice with lattice
constant $a$ and we take the average spin alignment of the magnetic sites to
be in the $z$ direction, which is also the direction of the applied magnetic
field. The nanoribbon is of finite width in the $y$ direction with $N$ atomic
rows (labeled as $n = 1,\cdots,N$) and it is infinite in the $x$ direction
($-\infty \Leftrightarrow \infty$). The position vector for each site is
given by $\mathbf{r} = a(m,n,0)$, where $m$ is an integer from $-\infty $ to
$\infty$, and $n$ is the row number with $n= 1,2, \cdots,N$ (see Figure
\ref{fig:squarelattice}).

The total Hamiltonian of the system is given by
\begin{equation}\label{dhimaltonian12} \hat{H}_{\textbf{Total}}=-\frac{1}{2}
\sum_{i,j} J_{i,j}\mathbf{S}_i\cdot\mathbf{S_j} -g\mu_B H_0 \sum_i S_i^Z
-\sum_i D_i (S_i^Z)^2,
\end{equation}
the first term is the Heisenberg NN exchange term, the second term is the
Zeeman energy term due to an applied field $H_0$, and the third term
represents the uniaxial anisotropy. The summations over $i$ and $j$ run over
all the sites, where the NN exchange $J_{ij}$ has the ``bulk" value $J$ when
either $i$ and $j$ are in the interior of the nanoribbon, and a modified
value $J_e$  when $i$ and $j$ are both at the edge of the nanoribbon (i.e.,
in row $n=1$ or $n=N$). Similarly, for the uniaxial anisotropy term, we
assume a value $D$ when the site $i$ is inside the nanoribbon, and a modified
value $D_e$ for sites at the edge of the nanoribbon.

To calculate the SWs for this system at low temperatures $T\ll T_{c}$ where
the spins are well aligned such that the thermal average of $S^z\simeq S$ for
each spin, we use the Holstein-Primakoff (HP) transformation and follow
similar procedures to \cite{Selim2011} to express the total Hamiltonian in terms of
boson operators. We arrive to the expression
\begin{equation}\label{df}
\hat{H}_{\textbf{Total}}= E_0+\hat{H}_s
\end{equation}
where the constant term $E_0$ is the energy of the ground state for the
ferromagnetic system given by
\begin{equation}\label{fgh}
 E_0=
 S^2\left(-\frac{1}{2}\sum_{i,j} J_{i,j}  -\sum_i D_i\right)-g\mu_B H_0 \sum_i S,
\end{equation}
and the operator term
$\hat{H}_s$ has the following quadratic form
\begin{equation}\label{rrrr2}
\hat{H}_s=-\frac{1}{2} S \sum_{i,j} J_{i,j}  \left(b_ib^{\dag}_j+b^{\dag}_ib_j- b^{\dag}_jb_j- b^{\dag}_ib_i \right) + \sum_i \left[g\mu_B H_0+(2S-1) D_i\right]b^{\dag}_ib_i
\end{equation}
where $b^{\dag}_j$ and $b_j$ are the creation and the annihilation boson
operators.

In order to diagonalize $\hat{H}_s$ and obtain the SW frequencies, we may
consider the time evolution of the creation and the annihilation operators
$b^{\dag}_j$ and $b_j$, as calculated in the Heisenberg picture in quantum
mechanics. In this case, the equation of motion
\cite{Bes2007,Liboff,Kantorovich2004,Roessler2009,HenrikBruus2004} (using
units with $\hbar=1$) for the annihilation operator $b_j$ is
\begin{align}
    \frac{d b_j(t)}{dt}&= i\left[H, b_j(t)\right] \nonumber\\
                       &= (g\mu_B H_0+(2S-1) D_j) b_j(t) -\frac{1}{2}S \sum_{i,j} J_{i,j} \left(b_i(t)- b_j(t)\right)  \label{eqmation}
\end{align}
where the commutation relation between $b^\dag_i$ and $b_j$ in
\begin{equation}\label{crp}
\left[b_i,b^\dag_j\right]=\delta_{ij},\hspace{20pt}  \left[b^\dag_j,b_i\right]=-\delta_{ij},\hspace{20pt} \left[b_i,b_j\right]=\left[b^\dag_j,b^\dag_i\right]= 0.
\end{equation}
was used, as well as the operator identity
$[AB,C]=A[B,C]+[A,C]B$.

The equation of motion for the creation operator $b_j^\dag$ is easily
obtained by taking the Hermitian conjugation of Equation~\eqref{eqmation},
giving
\begin{equation}\label{eqmation2}
i \frac{d b_j^\dag(t)}{dt}=  -(g\mu_B H_0+(2S-1) D_j) b_j^\dag(t) +\frac{1}{2}S \sum_{i,j} J_{i,j}  \left(b_i^\dag(t)- b_j^\dag(t)\right).
\end{equation}

The dispersion relations of the SWs (i.e., the energy or frequency versus
wavevector) can now be obtained by solving the above operator equations of
motion. The SW energy is related to the SW frequency using $E=\hbar \omega$,
and a Fourier transform for the operators from the time representation to the
frequency representation is made:
\begin{align}
b_j(x,t)&=\int_{-\infty}^{\infty} b_j(x,\omega)e^{-i\omega t}d\omega,  \nonumber\\
b_j^\dag(x,t)&=\int_{-\infty}^{\infty} b_j^\dag(x,\omega)e^{-i\omega t}d\omega. \label{frequency}
\end{align}
On substituting Equation~\eqref{frequency} in Equations \ref{eqmation} and
\ref{eqmation2}, we get
\begin{align}
    &\left(\omega  -  (g\mu_B H_0+(2S-1) D_j) -\frac{1}{2}S \sum_{i} J_{i,j} \right) b_j(\omega)+\frac{1}{2}S \sum_{i} J_{i,j} b_i(\omega)=0, \nonumber\\
    &\left(\omega  +  (g\mu_B H_0+(2S-1) D_j) +\frac{1}{2}S \sum_{i} J_{i,j} \right) b_j^\dag(\omega)-\frac{1}{2}S \sum_{i} J_{i,j}   b_i^\dag(\omega)=0. \label{operator2}
\end{align}

Since the nanoribbon extends to $\pm\infty$ in the $x$ direction, we may
introduce a 1D Fourier transform to wavevector $q_x$ along the $x$ direction
for the boson operators $b^{\dag}_j$ and $b_j$ as follows:
\begin{align}
b_j(x,\omega)&= \frac{1}{\sqrt{N_0}} \sum_{q_x} b_{n}(q_x,\omega) e^{i q_x ma}, \nonumber\\
b_j^\dag(x,\omega)&=\frac{1}{\sqrt{N_0}} \sum_{q'_x} b_{n'}^\dag(q'_x,\omega) e^{i q'_x m'a}, \label{operator1}
\end{align}
where $N_0$ is the (macroscopically large) number of spin sites in any row.
The transformed operators obey the following commutation relations:
\begin{equation}\label{}
\left[b_{n}(q_x,\omega),b^\dag_{n}(q'_x,\omega)\right]=\delta_{q_xq'_x}.
\end{equation}
By substituting Equation~\eqref{operator1} into Equation~\eqref{operator2} and
rewriting the summations, we get the following set of coupled equations:
\begin{align}
     &\left(\omega  -  (g\mu_B H_0+(2S-1) D_{s}) -\frac{1}{2}S (2J_e+J) +\frac{1}{2}S J_e \gamma(q_x)\right) b_{N}(q_x,\omega) \nonumber\\
     &\phantom{hjfhshfsddfjh} +\frac{1}{2}S J b_{N-1}(q_x,\omega)=0 \phantom{hjfhshfsddfjh ddsdds} \text{for }\, n=N \nonumber\\
     &\left(\omega  -  (g\mu_B H_0+(2S-1) D) -\frac{1}{2}S (4J) +\frac{1}{2}S J \gamma(q_x)\right) b_{n}(q_x,\omega) \label{efth} \\
     & \phantom{hjfhshfsddfjh}+\frac{1}{2}SJ(b_{n+1}(q_x,\omega) +b_{n-1}(q_x,\omega))=0  \phantom{hjfhshfsddfjh} \text{for }\, N>n> 1 \nonumber\\
     &\left(\omega  -  (g\mu_B H_0+(2S-1) D_{s'}) -\frac{1}{2}S (2J_{s'}+J) +\frac{1}{2}S J_{s} \gamma(q_x)\right) b_{1}(q_x,\omega)  \nonumber\\
     & \phantom{hjfhshfsddfjh}+\frac{1}{2}SJ b_{2}(q_x,\omega)=0  \phantom{hjfhshfsddfjh ddsdds} \text{for }\, n=1.   \nonumber
\end{align}
The first and the third equations refer to the edges $n=N$ and $n=1$ for the
nanoribbon system, and we have defined $\gamma(q_x)=2\cos(q_xa)$. Similar
results can be found for the equations involving the creation operator
$b_j^\dag$.

The above coupled equations can conveniently  be written in matrix form as
\begin{align}
    &(-\Omega I+A)b=0, \nonumber\\
    &(\Omega I+A)b^\dag=0,
\end{align}
where $b$ and $b^\dag$ are $N\times 1$ column matrices whose elements are the
boson operators $b_{n}(q_x,\omega)$ and $b_{n}^\dag(q_x,\omega)$. Also
$\Omega = \omega/SJ$ is a dimensionless frequency. The second equation is
redundant in that it does not give rise to any new physical modes, so we will
therefore ignore it.  Here $I$ is the $N\times N$ identity matrix and $A$ is
the following tridiagonal $N\times N$ matrix:

\begin{equation}A=\left(
    \begin{array}{cccccccc}
    a_s& -1& 0& 0& \cdots&  &   &   \\
    -1 &  a&-1& 0& \cdots&  &   &   \\
    0  & -1& a&-1& \cdots&  &   &   \\
     \vdots  &   \vdots&     \vdots&     \vdots& & \cdots &\cdots   &\cdots   \\
    & & & & \cdots& a &-1   &0   \\
    & & & & \cdots& -1&a   &-1   \\
    & & & & \cdots&  0& -1  & a_s
\end{array}\right). \label{amtrixs}
\end{equation}
The following dimensionless quantities have been defined:
\begin{align}
a_s&=\frac{2(g\mu_B H_0+(2S-1) D_e) -S(2J_e+J)+SJ_e \gamma(q_x)}{SJ},\\
a&=\frac{2(g\mu_B H_0+(2S-1) D) -S(4J)+SJ \gamma(q_x)}{SJ}. \nonumber
\end{align}
It is convenient to denote a new matrix by $A'=-\Omega I+A$, so

\begin{equation}A'=\left(
    \begin{array}{cccccccc}
    a'_s& -1& 0& 0& \cdots&  &   &   \\
    -1 &  a'&-1& 0& \cdots&  &   &   \\
    0  & -1& a'&-1& \cdots&  &   &   \\
     \vdots  &   \vdots&     \vdots&     \vdots& & \cdots &\cdots   &\cdots   \\
    & & & & \cdots& a' &-1   &0   \\
    & & & & \cdots& -1&a'   &-1   \\
    & & & & \cdots&  0& -1  & a'_s
\end{array}\right),
\end{equation}
where
\begin{align}
a'_s&=\frac{-\omega+2(g\mu_B H_0+(2S-1) D_e) -S(2J_e+J)+SJ_e \gamma(q_x)}{SJ}, \nonumber\\
a'&=\frac{-\omega+2(g\mu_B H_0+(2S-1) D) -S(4J)+SJ \gamma(q_x)}{SJ}. \label{eqas}
\end{align}
The new tridiagonal matrix $A'$ may be separated into two terms, following
the approach in \cite{PhysRev.185.720,Cottam1976,RefWorks:21}, as
\begin{equation}
    A'=A_0+\mathbf{\Delta},
\end{equation}
where
\begin{equation}A_0=\left(
    \begin{array}{cccccccc}
    a'& -1& 0& 0& \cdots&  &   &   \\
    -1 &  a'&-1& 0& \cdots&  &   &   \\
    0  & -1& a'&-1& \cdots&  &   &   \\
     \vdots  &   \vdots&     \vdots&     \vdots& & \cdots &\cdots   &\cdots   \\
    & & & & \cdots& a' &-1   &0   \\
    & & & & \cdots& -1&a'   &-1   \\
    & & & & \cdots&  0& -1  & a'
\end{array}\right),
\end{equation}

\begin{equation}\mathbf{\Delta}=\left(
    \begin{array}{cccccccc}
    \Delta& 0& 0& 0& \cdots&  &   &   \\
    0 &  0&0& 0& \cdots&  &   &   \\
    0  & 0& 0&0& \cdots&  &   &   \\
     \vdots  &   \vdots&     \vdots&     \vdots& & \cdots &\cdots   &\cdots   \\
    & & & & \cdots& 0 &0   &0   \\
    & & & & \cdots& 0& 0   &0   \\
    & & & & \cdots&  0& 0  &    \Delta
\end{array}\right),
\end{equation}
and the element $\Delta=a'_s-a'$. In this way all the edge properties have
been separated into the matrix $\mathbf{\Delta}$. The inverse of a
finite-dimensional tridiagonal matrix with constant diagonal elements such as
$A_0$ is well known \cite{Cottam1980,PhysRev.185.720} and can be expressed as
\begin{equation}\label{inv:1}
(A_0^{-1})_{ij}=\frac{x^{i+j}-x^{|i-j|}+x^{2N+2-(i+j)}-x^{2N+2-|i-j|}}{\left(1-x^{2N+2}\right)\left(x-x^{-1}\right)}.
\end{equation}
Here $x$ is a complex variable defined such that $|x|\leq 1$ and
$x+x^{-1}=a'$.

On noting that $A'=(A_0+\mathbf{\Delta})=A_0(1+A_0^{-1}\mathbf{\Delta})$, the
dispersion relations are obtained by the condition
\cite{Cottam2004,RefWorks:21} that $\det A'=0$, which implies
\begin{eqnarray} \label{condition}
\det(I+A_0^{-1}\mathbf{\Delta})    &=&0.
\end{eqnarray}

Using the previous equations, the matrix $M =(I+A_0^{-1}\mathbf{\Delta})$ can
next be written in a partitioned form \cite{algebra}:
\begin{equation}\label{f}
  M= \left(%
\begin{array}{c|c|c}
  M_{1,1} &  0&  M_{1,N} \\ \hline
  M_{2,1} &  &  M_{2,N} \\
  \vdots & I &  \vdots  \\
  M_{N-1,1} &  &  M_{N-1,N} \\\hline
 M_{N,1}&  0& M_{N,N}
\end{array}%
\right),
\end{equation}
where the nonzero elements of $M$ can be written as
\begin{equation}\label{m:1}
    M_{i,j}=\delta_{i,j}+ \delta_{1,j}(A_0^{-1})_{1,j}\triangle
    +\delta_{N,j}(A_0^{-1})_{i,N}\triangle,
\end{equation}
and the determinant of $M$, which is required for Equation~\eqref{condition}, can
be calculated to give
\begin{eqnarray}
\det(M) &=& (M_{1,1})^2-(M_{1,N})^2  \nonumber\\
&=& \left(1+\frac{x^{2}+x^{2N}-x^{2N+2}-1}{\left(1-x^{2N+2}\right)\left(x-x^{-1}\right)}\Delta\right)^2 \\
&& - \left(\frac{2x^{N+1}-x^{N-1}-x^{N+3}}{\left(1-x^{2N+2}\right)\left(x-x^{-1}\right)}\Delta \right)^2. \nonumber
\end{eqnarray}
After some more algebraic steps the condition for $\det(M)=0$ can be written
as
\begin{eqnarray}
&&\left[\left(1-x^{2N+2}\right)\left(x-x^{-1}\right)+\left(x^{2}+x^{2N}-x^{2N+2}-1\right)\Delta\right]  \nonumber\\
  &&+\eta\left(2x^{N+1}-x^{N-1}-x^{N+3}\right)\Delta=0, \label{dispersion}
\end{eqnarray}
where $\eta= \pm 1$. This result is formally similar to an expression
obtained in the study of finite thickness ferromagnetic slabs
\cite{Cottam1980}. Also, it follows by analogy with previous work
\cite{Cottam1976,Cottam1980} that the solutions with $|x|=1$ correspond to
the area modes (those propagating across the width of the stripe) while those
with $|x|<1$ correspond to the localized edge modes.

Before presenting a general analysis of Equation~\eqref{dispersion} for the
SW frequencies in Section \ref{sect3}, we next examine some special cases.

\subsection{Special case of $N\rightarrow\infty$}
It is of interest to study the behavior of the model in the special case
when the ribbon (or stripe) becomes very wide. This is the limit of $N\rightarrow\infty$ in
Equation~\eqref{dispersion}. Since $|x|<1$ for edge modes, the terms of
order $x^N\rightarrow0$ as $N\rightarrow\infty$, giving for these modes
\begin{eqnarray}\label{con1234}
   && (1+x\Delta)=0\hspace{30pt} \Rightarrow x=-\frac{1}{\Delta}.
\end{eqnarray}
This is the same formal expression as obtained in the case of a semi-infinite
Heisenberg ferromagnet \cite{Cottam1976, Cottam1980} when
$N\rightarrow\infty$.

For the edge mode localization condition $|x|<1$ to be satisfied
in this special case, Equation \ref{con1234} implies $|\Delta|>1$, which
gives two possibilities. The first one is $\Delta>1$ and the second one is
$\Delta<-1$. From the definitions of $a'_s$ and $a'$
we have
\begin{align*}
      \Delta  &=\frac{2(2S-1)(D_e-D) -S(2J_e-3J)+S(J_e-J)
        \gamma(q_x)}{SJ},
\end{align*}
which depends on the physical parameters $D_e,\, D, \, S,\, J_e$ and $J$ of the ferromagnetic stripe and on the wavevector component $q_x$.
Since $\gamma(q_x)=2\cos(q_xa)$ has its
maximum when $\cos(q_xa)=1$ and its minimum when $\cos(q_xa)=-1$, it follows
that the minimum value $\Delta_{min}$ and the maximum value $\Delta_{max}$
for $\Delta$ correspond to
\begin{align*}
\Delta_{max/min}&=\frac{2(2S-1)(D_e-D) -S(2J_e-3J)\pm 2S(J_e-J)}{SJ}
\end{align*}
In simple cases we might have $D_e=D$ i.e., the edge perturbation to the anisotropy is negligible. Then, denoting
the ratio between the edge exchange and area exchange by
$r=J_e/J>0$, we have
\begin{align*}
\Delta_{max}   &=1\hspace{30pt} \Delta_{min} =-4r+5.
\end{align*}
The two cases $\Delta>1$ and $\Delta<-1$ give the following ranges for
$r$:
\begin{eqnarray}
r<\frac{5}{4} \hspace{50pt}  r>\frac{6}{4}.   \label{edge22}
\end{eqnarray}
These are the ranges of the ratio exchange $r=J_e/J$ for which the edge modes exist at some $q_x$ value in this
special case ($N\rightarrow\infty$) for the square lattice.

The conditions are modified if edge perturbations in the anisotropy are included.

\subsection{Case of large finite $N$}
Another interesting case is when $N$ becomes sufficiently large that
the two solution $x^+$ and $x^-$ for $\eta= \pm 1$ of
Equations~\eqref{dispersion} can be obtained by an iterative approach used in
reference \cite{Cottam1980}. Since $|x|<1$ for edge modes, all terms of order $x^N$
in Equations~\eqref{dispersion} are small for sufficiently large $N$ and
the two solution $x^+$ and $x^-$  become closer to the solution $x_0 =
-\Delta^{-1}$ for the special case of $N\rightarrow\infty$.
\begin{figure}[h!]
\centering
\includegraphics[scale=1]{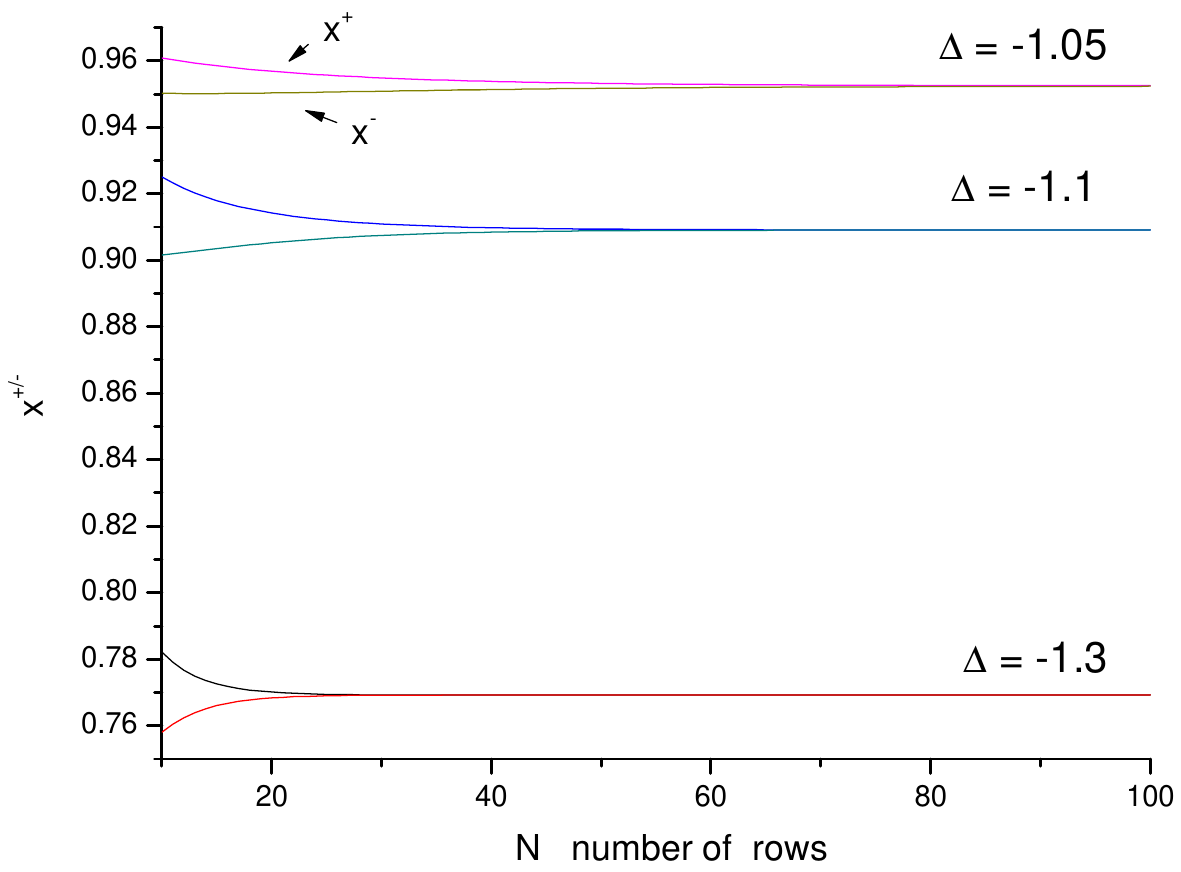}
\caption{Calculated values of $x^+$ and $x^-$ for several values of $\Delta <-1$ and for $N$ from 10 to 100.} \label{fig:largn3}
\end{figure}

To use a first order iteration for Equations~\eqref{dispersion} we must
rewrite them in the forms $x^{\pm}=F^{\pm}(x_0)$.
We get the following two first order iteration approximate solutions
\begin{equation}\label{f}
    x^{\pm}=\frac{-1}{x_{ap/am}}
\end{equation}
where $x_{ap}$ is equal to
\[
x^{2N+3}_0+\Delta x^{2N+2}_0-x^{2N+1}_0-\Delta x^{2N}_0+\Delta x^{N+3}_0-2\Delta x^{N+1}_0+\Delta x^{N-1}_0
 +\Delta
\]
and $x_{am}$ is equal to
\[
x^{2N+3}_0+\Delta x^{2N+2}_0-x^{2N+1}_0-\Delta x^{2N}_0- \Delta x^{N+3}_0+2\Delta x^{N+1}_0-\Delta x^{N-1}_0
+\Delta.
\]
This approximation is valid provided  $|\Delta|>1$. Figure \ref{fig:largn3}
shows calculated values of $x^+$ and $x^-$ using illustrative values of
$\Delta$ chosen as $-1.05$, $-1.1$ and
$-1.3$ when $N$ has values from 10 to 100. The iterative approach is found to work well for $N > 10$ in this case.

\section{Numerical calculations}\label{sect3}
More generally, the dispersion relations can be obtained by solving Equation
\eqref{dispersion} using a numerical calculation for any finite $N$. The number
of rows $N$ and the value of $\Delta$ are first substituted in the Equations
\eqref{dispersion}, and then the polynomial equations are solved for $x$
which can be used to obtain the dispersion relations. Since the solutions for
$x$ may have complex roots, one of the ways to solve such equations is to use
Laguerre's method for finding the roots\cite{RefWorks:27,RefWorks:28} of
polynomials. By rearranging Equations \eqref{dispersion} we have:
\begin{eqnarray} \nonumber
\text{First polynomial}  \phantom{ssssssssssssssssssss}&&  \\
\nonumber
   x^{2N+4}+\Delta x^{2N+3}-x^{2N+2}-\Delta x^{2N+1}- \Delta x^{N+4}+2\Delta x^{N+2}-\Delta x^{N}\\
 -\Delta x^{3}- x^{2}+\Delta x+1=0 \label{fp}\\ \nonumber
\text{Second polynomial} \phantom{ssssssssssssssssssss} &&
\\ \nonumber
 x^{2N+4}+\Delta x^{2N+3}-x^{2N+2}-\Delta x^{2N+1}+ \Delta x^{N+4}-2\Delta x^{N+2}+\Delta x^{N} \label{sp}\\
 -\Delta x^{3}- x^{2}+\Delta x+1=0
 \end{eqnarray}

Both polynomials are of degree $2N+4$, and they are applicable for all $N$ equal to 3 and above.
We note that there is a special case when $N=3$, since the two power indices $2N+1$ and
$N+4$ become equal to 7. The obtained values for $x$ must satisfy the conditions for physical SW modes mentioned earlier.
The edge SW modes are
localized on the edge and their amplitudes decay exponentially inside the nanoribbon. This
requires that $x$ must be real and less than 1 for edge modes. The area modes
are oscillating waves inside the nanoribbon, and so
\begin{eqnarray}
  && x \in R \text{ and } |x|<1  \text{ for edge modes} \nonumber\\
  && x=e^{iq_ym} \text{ and } |x|\leq 1 \text{ for area modes} \label{cond}
\end{eqnarray}

In the previous section the ranges for $r$ for edge modes to exist were obtained algebraically for the special cases
$N\rightarrow\infty$ and large $N$. We can use numerical calculations to obtain the ranges of $\Delta$ for smaller $N$
that satisfies the conditions for which both edge
modes and area modes exist.

For that purpose, a Fortran program was written to solve the two polynomials using Laguerre's method where two subroutines \verb"zroots" and
\verb"laguer" are adapted from \cite{RefWorks:27}.

\begin{figure}[hp]
\centering
\includegraphics[scale=1]{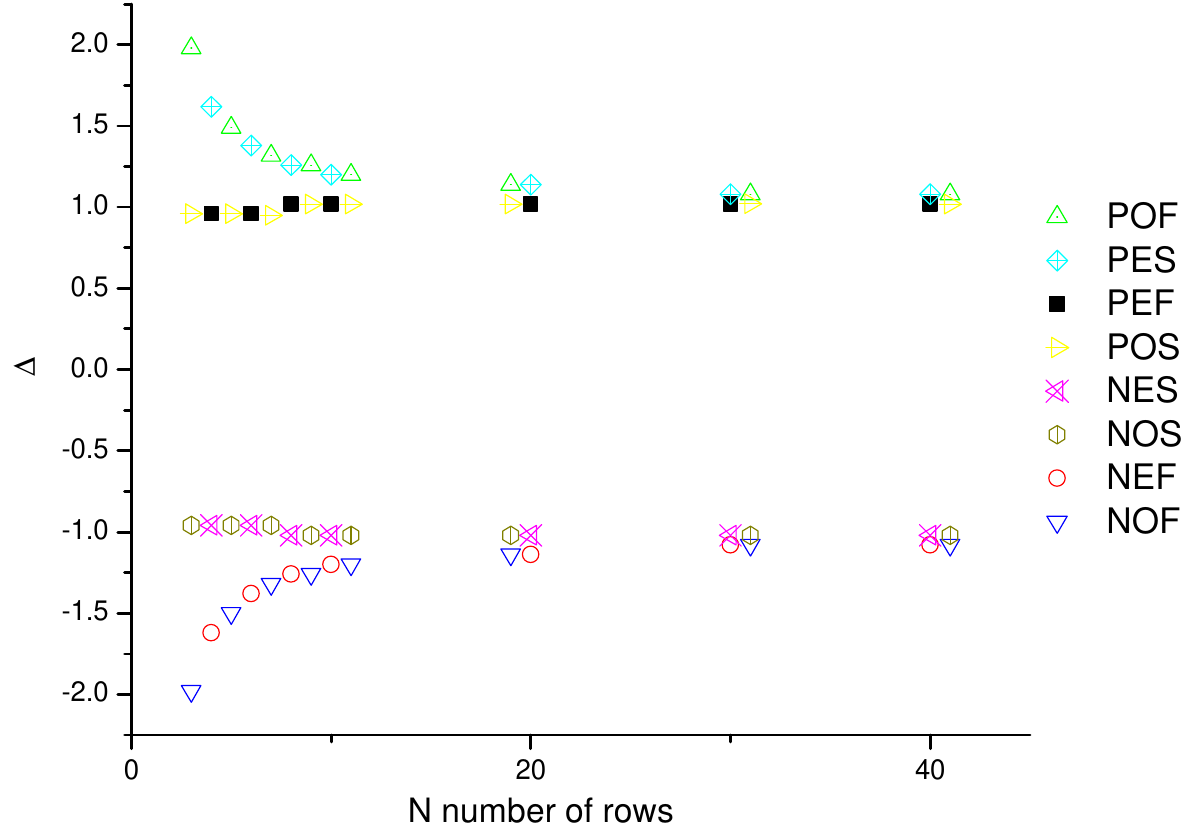}
\caption{The values for minimum positive (P) and maximum negative (N) of $\Delta$ for even (E) and odd (O) rows
number $N$ that
satisfies edge modes \eqref{cond}, are computed from the first polynomial (F) \ref{fp}
and from the second polynomial (S) \ref{sp}.} \label{fig:223}
\vspace{20pt}
\centering
\includegraphics[scale=1]{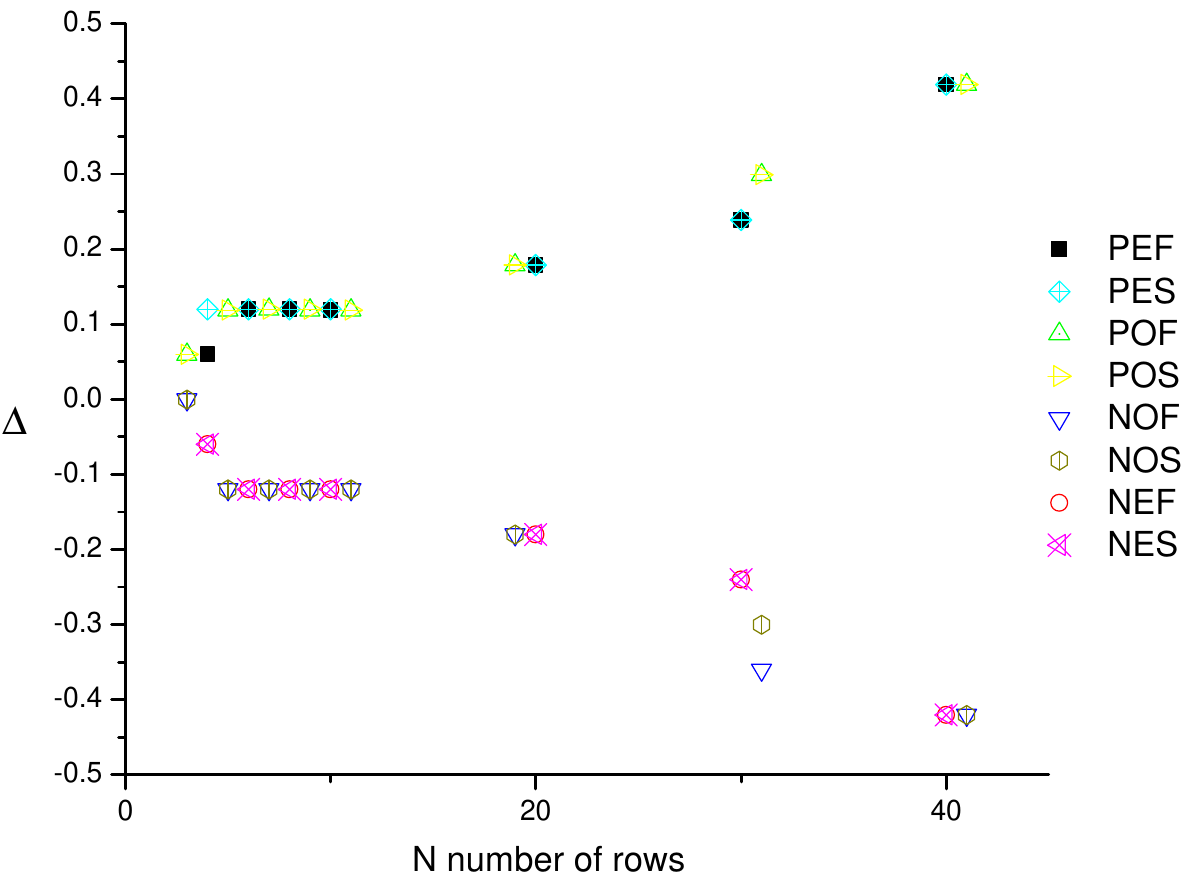}
\caption{The values for minimum positive (P) and maximum negative (N) of $\Delta$ for even (E) and odd (O) rows
number $N$ that
satisfies area modes \eqref{cond} are computed from the first polynomial (F) \ref{fp}
and from the second polynomial (S) \ref{sp}.} \label{fig:224}
\end{figure}

The values for minimum positive (P) and maximum negative (N) of $\Delta$ for
even (E) and odd (O) rows number $N$ that satisfy \eqref{cond} are computed
from the first polynomial (F) \ref{fp} and from the second polynomial (S)
\ref{sp}  and displayed in figure \ref{fig:223} for edge modes and in figure
\ref{fig:224} for area modes.

Figure \ref{fig:223} shows the behavior of minimum positive for $\Delta$ that
satisfies the conditions \eqref{cond} for the existence of edge modes. It is
clear from the figure, that in the range of rows number $N\leq20$, the
minimum positive of both odd rows number of first polynomial \ref{fp} and
even rows number for second polynomial \ref{sp} are approximately the same
and are exponentially decaying to a nearly constant value of 1.02. In the
same rows number range, the minimum positive of even rows for first
polynomial \ref{fp} and of odd rows for second polynomial \ref{sp} is nearly
constant and equal to 0.95. After $N$ equal to 20, the minimum positive of
both polynomials is independent of stripe (ribbon) width, (i.e. rows number),
and it is also independent of rows number parity, whether even or odd. As the
rows number increases, the minimum positive is convergent to an approximate
constant value of 1.02.

The same Figure \ref{fig:223} displays the maximum negative of $\Delta$ for
first polynomial \ref{fp} and second polynomial \ref{sp}. Here, the maximum
negative shows different behavior from that of the above minimum positive as
it is independent of rows number parity for both polynomial. In the range of
rows number $N\leq20$, as in case of minimum positive, the maximum negative
of first polynomial \ref{fp} is nearly constant and equal to -0.96, while the
maximum negative of second polynomial \ref{sp} exponentially increases to a
nearly constant value of 1.02. As in the case of the minimum positive, after
$N$ equal to 20, the maximum negative of both polynomials is independent of
stripe (ribbon) width (i.e. rows number), and rows number parity, whether
even or odd. As $N$ increases, the maximum negative value is convergent to
approximately constant value equal to -1.02.

The conclusion from Figure \ref{fig:223} is that edge modes, in small rows
number $N\leq20$, are dependent on both the stripe width and the rows number
parity.  This is an indication for the interaction between the two edges in
the small range of rows number. As $N$ increases above 20, the edge modes
become independent on both the stripe width and the rows number parity.  In
this case both minimum positive and maximum negative of both polynomials are
stripe width and rows number parity independent. This is an indication for
disappearing of the interaction between the two edges after $N=20$. That
behavior agrees with result for the special case of $N$ become large, as
discussed above. Also, it is noted in the range of $N$ larger than 20, that
the difference between minimum positive and maximum negative is nearly
constant and independent of the stripe width and rows number parity.

Figure \ref{fig:224} shows the behavior of minimum positive and maximum
negative for $\Delta$ that satisfies the conditions \eqref{cond} for area
modes. It is clear that both minimum positive and maximum negative are
independent of rows number parity. While they depend on the stripe width, in
the range of rows number from 5 to 10 the two values are constant. After $N
=10$, the value of minimum positive is increasing linearly  with the stripe
width, while the value of maximum negative is decreasing linearly with the
stripe width. The difference between minimum positive and maximum negative is
increasing as the stripe width increases and it is mostly independent on the
rows number parity.

To obtain the dispersion relations for the above system, a Fortran program
was written to solving the first polynomial
\ref{fp} and the second polynomial \ref{sp} by Laguerre's method using two
subroutines \verb"zroots" and \verb"laguer" adopted from \cite{RefWorks:27},
which are used before. The values of physical parameters for calculating
these dispersion relations are chosen as follow:  $S=1,\, J=1,\, D=D_e=0$ and
$ g\mu_B H_0=0.3J$. The chosen value for the ratio between the edge exchange
and area exchange is equal to $r=J_e/J=0.04$, which is satisfies the
existence condition \ref{edge22} for edge mode. The chosen values of $q_xa$
are run from 0 to $\pi$ corresponding to the first Brillouin zone center and
boundary respectively.

\section{SW dispersion relations}
\begin{figure}[hp!]
\centering
\includegraphics[scale=1]{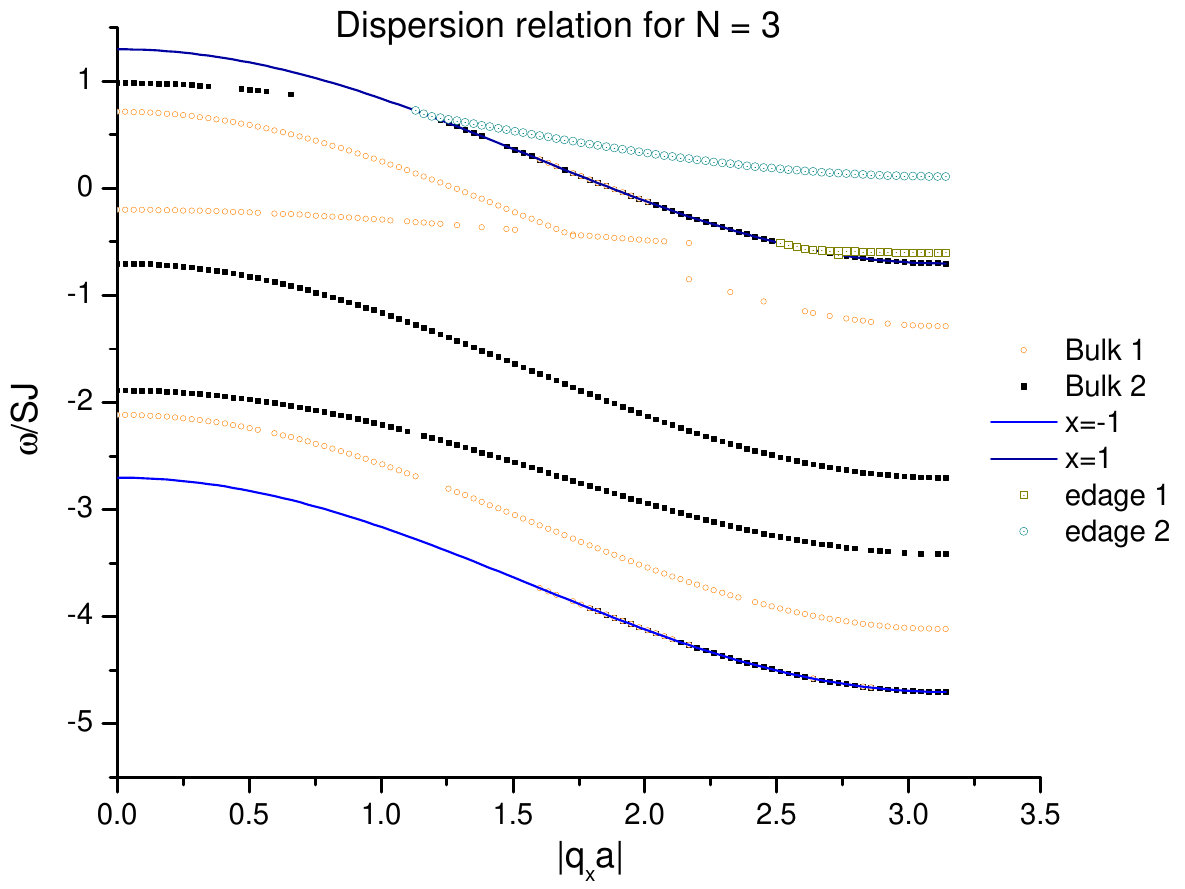}
\caption{Area and edge spin waves modes (in units of $SJ$) plotted against the wavevector $q_xa$ for stripe with width $N=3$, where $x=1$ and $x=-1$ are
the upper and lower boundary for area modes.} \label{fig:n3}
\vspace{20pt}
\centering
\includegraphics[scale=1]{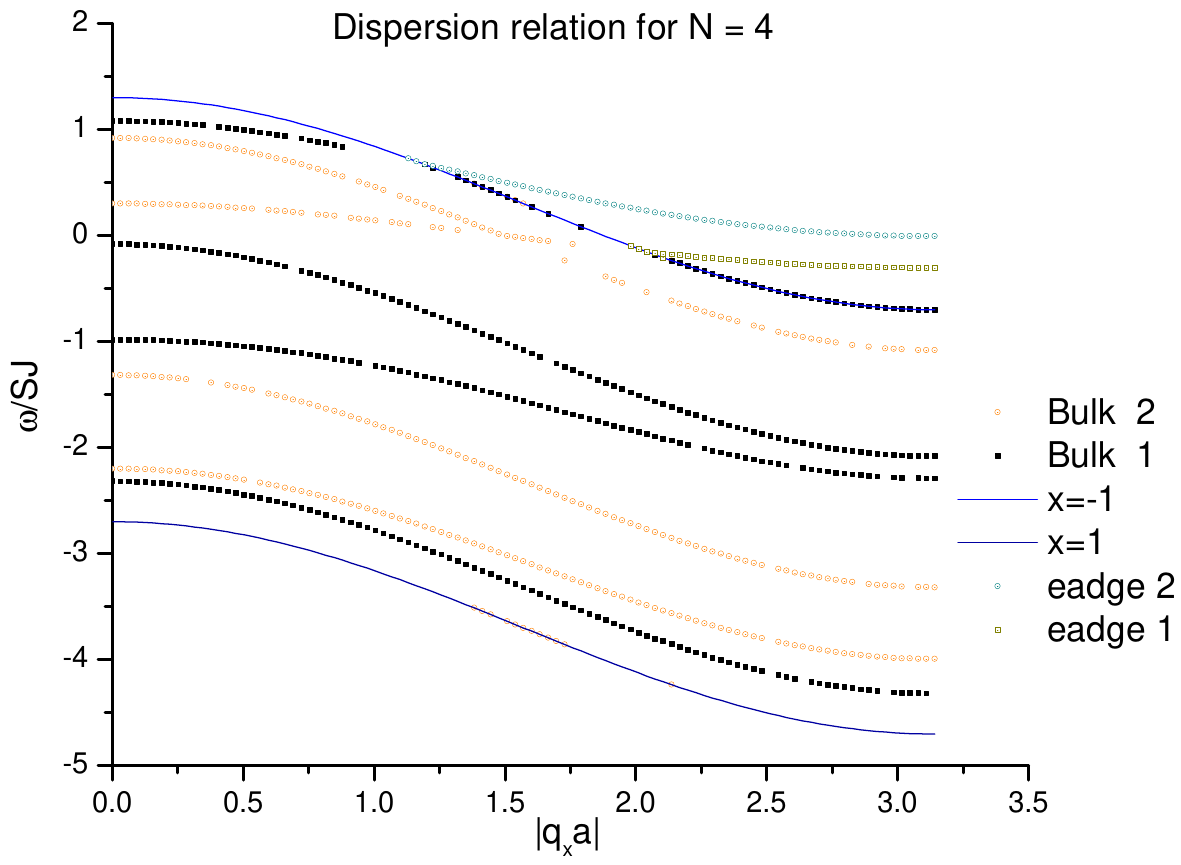}
\caption{Area and edge spin waves modes (in units of $SJ$) plotted against the wavevector $q_xa$ for stripe with width $N=4$, where $x=1$ and $x=-1$ are
the upper and lower boundary for area modes.} \label{fig:n4}
\end{figure}

\begin{figure}[hp!]
\centering
\includegraphics[scale=1]{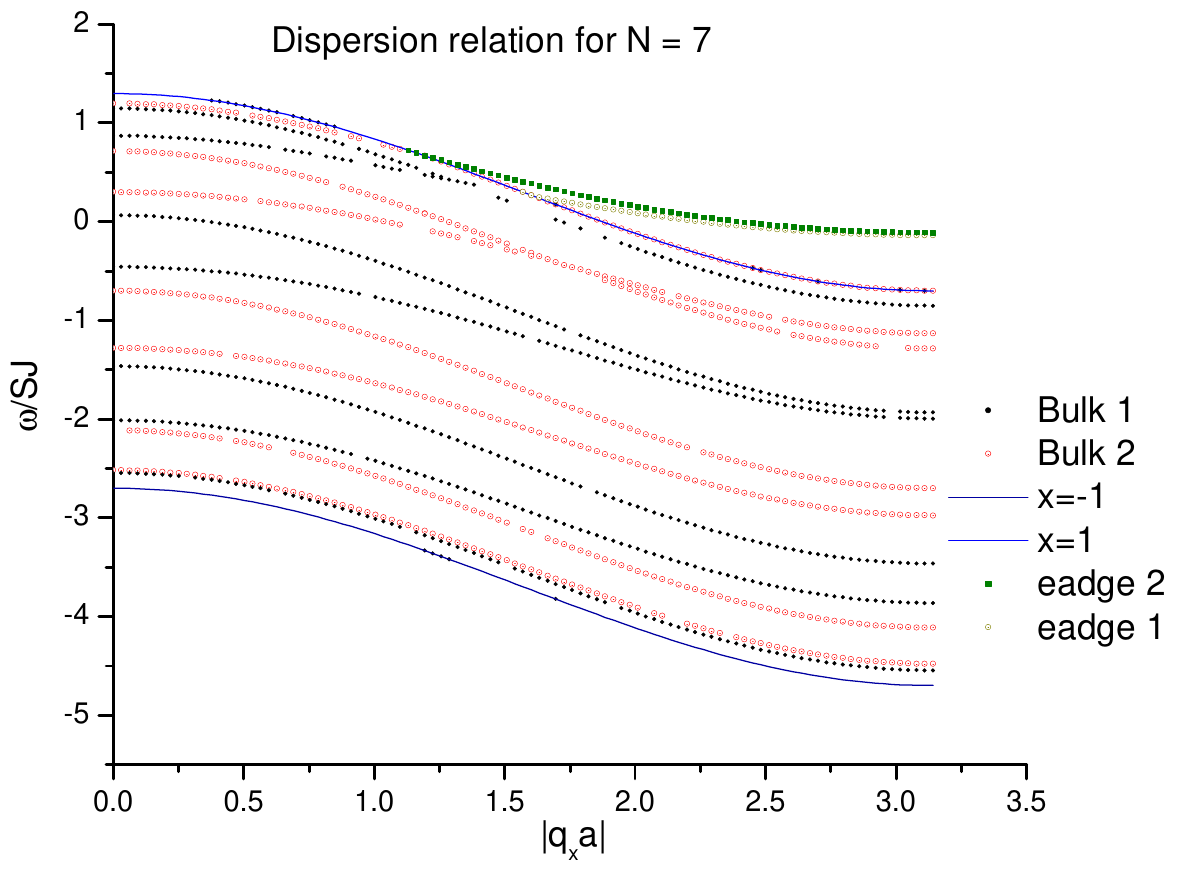}
\caption{Area and edge spin waves modes (in units of $SJ$) plotted against the wavevector $q_xa$ for stripe with width $N=7$, where $x=1$ and $x=-1$ are
the upper and lower boundary for area modes.} \label{fig:n7}
\vspace{20pt}
\centering
\includegraphics[scale=1]{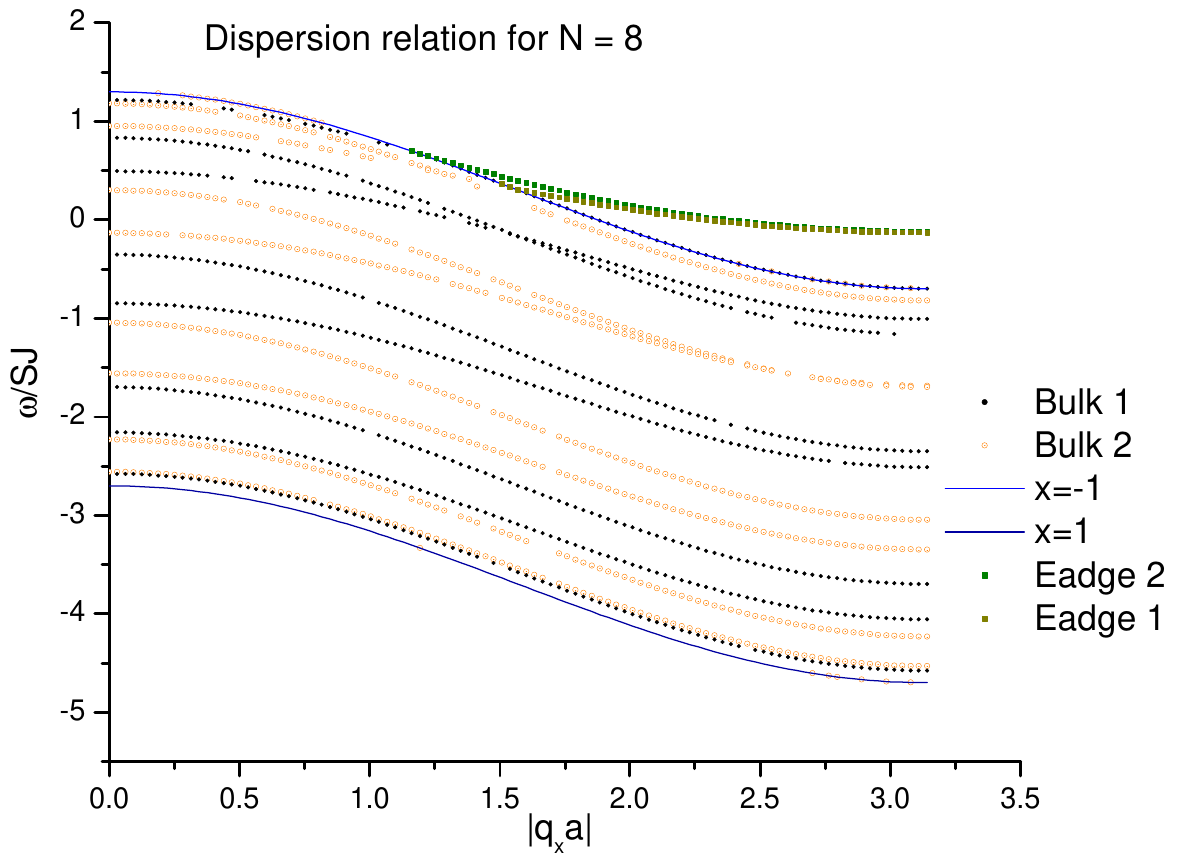}
\caption{Area and edge spin waves modes (in units of $SJ$) plotted against the wavevector $q_xa$ for stripe with width $N=8$, where $x=1$ and $x=-1$ are
the upper and lower boundary for area modes.} \label{fig:n8}
\end{figure}

The numerical results for calculating the dispersion relations of 2D square
lattice using the above algorithm and physical parameters are displayed in
Figures \ref{fig:n3}-\ref{fig:n8}.  These figures show plots of SW
frequency in terms of the dimensionless quantity $\omega/SJ$ as a function of the dimensionless
wavevector $q_xa$  for various numbers of rows. The two polynomials \ref{fp} and
\ref{sp} have even power in $x$, and therefore the area and edge modes are
symmetric about $\omega/SJ=0$. As a result, we have chosen to show only the
positive frequency branches.

All the figures display similar general features for ferromagnetic 2D
stripes.

Since every polynomial gives $N$ area modes, the number of area modes are
equal to twice the number of rows $N$. These area modes are upper bounded by
the frequency obtained when the value of $x$ is equal to $=1$, and lower
bounded by the frequency obtained when the value of $x$ is equal to $=-1$, as
shown in the Figures. In the upper of area modes, some spin wave frequencies
cross each other.  As $N$ increases, the number of areas modes inside their
boundary increases twofold, which then merge into an areas modes continuum.

The figures show that in all cases that there are two optic edge modes
appearing above the area modes region, these two edge modes look like
extension for their counterpart area modes. As $N$ increases, the difference
between the two edge modes is decreasing which is seen too for their
counterpart area modes.
\section{Discussion}
The dispersions relations of area and edges spin waves, and the effect of the
stripe width on them for 2D ferromagnetic square lattice stripes, have been
studied using the tridiagonal method. The result shows the same unexpected
feature: the area and edge spin waves only exist in optic modes.  This
behavior is also seen in 2D Heisenberg antiferromagnetic square lattice
experimental and theoretically
\cite{PhysRevLett.105.247001,PhysRevLett.105.157006,PhysRevLett.86.5377}, the
absence of the acoustic modes could be explained that the square lattice
support only optics modes, which need more studies to be completely
understood.

Our conclusion, that the unexpected behavior of spin waves in the 2D square
lattice of existence in only optic modes if included in the HTS theories
may lead to an explanation for HTS. Since it is known that HTS is linked to
2D square antiferromagnetic lattice, and  we expect that the Optic spin wave
could mediate the electrons using their spin degree of freedom and Pauli
exclusion principle for the formation of cooper pair with much less energy
than cooper pair created by phonon mediated electrons using their electric
charge
\cite{JPSJ.57.1544,0256-307X-18-10-332,Guo200179,G.M.Zhang,Dahm2009,Zhao2009}.

\begin{acknowledgments}
This research has been supported by the Egyptian Ministry of Higher Education
and Scientific Research (MZA).
\end{acknowledgments}

\bibliography{xbib2}

\end{document}